\documentclass[12pt]{iopart}

\usepackage{iopams}  

\usepackage{graphicx}
\usepackage{epstopdf}
\usepackage{color}

\begin{document}

\title[$^{31}$P NMR studies of the spin-$\frac52$ triangle lattice antiferromagnet  Na$_3$Fe(PO$_4$)$_2$]{Static and dynamic magnetic properties of the spin-$\frac52$ triangle lattice antiferromagnet Na$_3$Fe(PO$_4$)$_2$ studied by $^{31}$P NMR}

\author{Devi V. Ambika$^{1,2}$, Qing-Ping Ding$^{1,2}$,  Sebin J. Sebastian$^3$, Ramesh Nath$^3$, and  Yuji Furukawa$^{1,2}$}
\address{$^1$ Ames National Laboratory, U.S. DOE,  Iowa State University, Ames, Iowa 50011, USA}
\address{$^2$ Department of Physics and Astronomy, Iowa State University, Ames, Iowa 50011, USA}
\address{$^3$ School of Physics, Indian Institute of Science Education and Research Thiruvananthapuram-695551, India}

\ead{furukawa@ameslab.gov}
\begin{abstract}
  $^{31}$P nuclear magnetic resonance (NMR) measurements have been carried out to investigate the magnetic properties and spin dynamics of Fe$^{3+}$ ($S$ = 5/2) spins in the two-dimensional triangular lattice (TL) compound Na$_3$Fe(PO$_4$)$_2$.    The temperature ($T$) dependence of nuclear spin-lattice relaxation rates ($1/T_1$) shows a clear peak around  N\'eel  temperature, $T_{\rm N} = 10.9$~K, corresponding to an antiferromagnetic (AFM) transition.    From the temparature dependence of NMR shift ($K$) above $T_{\rm N}$,   an exchange coupling between Fe$^{3+}$ spins was estimated to be $J/k_{\rm B}\simeq 1.9$~K using the spin-5/2 Heisenberg isotropic-TL model.     The temperature dependence of $1/T_1T$ divided by the magnetic susceptibility ($\chi$),  $1/T_1T\chi$, above $T_{\rm N}$ proves the AFM nature of spin fluctuations below $\sim$ 50 K in the paramagnetic state.     In the magnetically ordered state below $T_{\rm N}$, the characteristic rectangular shape of the NMR spectra is observed, indicative of a commensurate AFM state in its ground state.     The strong temperature dependence of 1/$T_1$ in the AFM state is well explained by the two-magnon (Raman) process of the spin waves in a 3D antiferromagnet with a spin-anisotropy energy gap of 5.7 K.     The temperature dependence of sublattice magnetization is also well reproduced by the spin waves.     Those results indicate that the magnetically ordered state of Na$_3$Fe(PO$_4$)$_2$ is a conventional 3D AFM state, and no obvious spin frustration effects were detected in its ground state.
\end{abstract}

\maketitle

\section{Introduction}
 
  Geometrically frustrated magnets have attracted great interest because the systems show peculiar properties such as the suppression of the magnetic ordering, unconventional spin states like the spin liquid ~\cite{Li}, and large fluctuations in the ground state.
   Initially, the ground state of two-dimensional (2D) triangular lattice (TL) antiferromagnets  was predicted to host a resonating valence bond state~\cite{Anderson}, and the long-range magnetic ordering in TL was identified to occur when the spins orient at 120$^\circ$ angle relative to neighboring spins~\cite{Huse, Jolicoeur}. 
   Later studies of novel compounds proved the existence of non-trivial ground states in such systems. 
    For example, the presence of a weak easy-axis anisotropy in $S \geq 3/2$ triangular lattice antiferromagnets leads to a collinear ordered state~\cite{Sen}.

\begin{figure}[tb]
\includegraphics[width=12 cm]{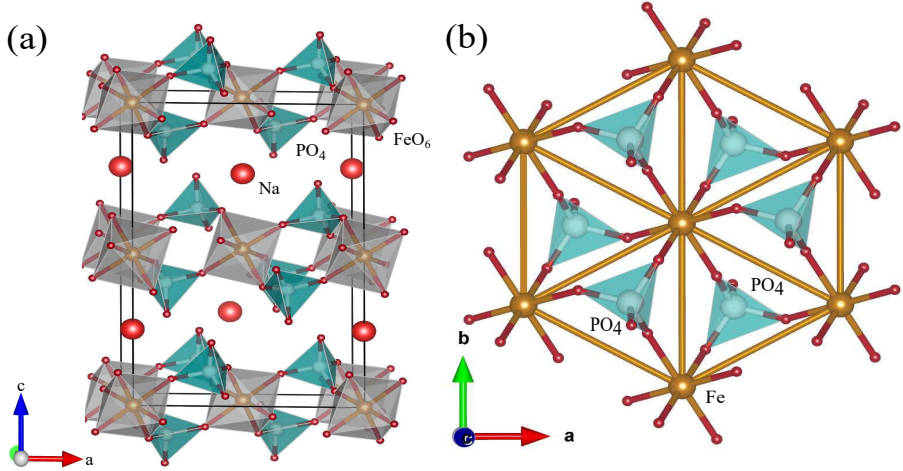} 
\caption{ (a) Crystal structure of Na$_3$Fe(PO$_4)_2$ showing corner-shared FeO$_6$ octahedra and PO$_4$ tetrahedra forming triangular layers. 
    Na atoms are located in-between the layers. (b) Fe$^{3+}$ ions connected via PO$_4$ tetrahedra forming the triangular lattice in the $ab$ plane.
}
\label{fig:structure}
\end{figure}

    Na$_3$Fe(PO$_4)_2$ is one of the series compounds of the family $AA^{\prime}M$(\textit{R}O$_4)_2$, (\textit{A} = Ba, K, Rb, Na; $A^\prime$ = $\rm {Na}_2$, $\rm {Ag}_2$; \textit{M} = Cr, Mn, Fe, Co, Ni, and Cu; and \textit{R} = P, V, As)~\cite{Tapp, Sebastian,Amuneke} which provides a good platform  to study systematically the magnetic properties of triangular lattice antiferromagnets  with $S \geq 3/2$. Depending upon the size of the non-magnetic spacers and exchange coupling, the systems can host either a magnetically ordered phase or a spin liquid phase.
    Na$_3$Fe(PO$_4)_2$ crystallizes in a monoclinic space group ($C2/c$) with the lattice parameters of  $a$ = 9.0714(2)~\AA, $b$ = 5.032(1)~\AA, and $c$ = 13.8683(3)~\AA~at room temperature ~\cite{Belkhiria,Morozov}.
    Fe$^{3+}$ ions ($S$ = 5/2) in distorted FeO$_6$ octahedra form an $S$ = 5/2 anisotropic TL via corner-sharing with PO$_4$ tetrahedra in the $ab$ plane as shown in Fig.~\ref{fig:structure}, where the triangles are slightly distorted due to the two different Fe-Fe distances of 5.182~\AA~and 5.028~\AA~\cite{Sebin}.
   Recently, Sebastian $et ~al.$  reported that Na$_3$Fe(PO$_4)_2$  exhibits an antiferromagnetic (AFM) ordering with a  N\'eel  temperature of $T_{\rm N} = 10.6$~K with a reduced Fe ordered magnetic moment of $\sim$ 1.52 $\mu_{\rm B}$ at 1.6 K in zero magnetic field~\cite{Sebin}. 
   The average exchange coupling between Fe$^{3+}$ spins was estimated to be $J/k_{\rm B}\simeq 1.8$~K by fitting magnetic susceptibility $\chi(T)$ data using the high-temperature series expansion for a spin-5/2 Heisenberg TL model.
    The authors also determined, from neutron diffraction measurements,  a collinear AFM structure  with a propagation vector $q$ = (1, 0, 0) where the Fe spins are coupled ferromagnetically along the $b$ axis or in the [100] plane~\cite{Sebin}.
However, the detailed magnetic properties, especially magnetic fluctuations, of  Na$_3$Fe(PO$_4)_2$ have still not been investigated. 

   Nuclear magnetic resonance (NMR) is a very powerful experimental tool to probe the static and dynamic magnetic properties of materials  from a microscopic point of view.
   In this paper, we report the results of $^{31}$P NMR measurement on the polycrystalline Na$_3$Fe(PO$_4)_2$ sample in the temperature range of $T =1.5 - 300$~K.
   The NMR spectra, nuclear spin-lattice relaxation rate 1/$T_1$ and nuclear spin-spin relaxation rate 1/$T_2$ measurements show the  AFM long-range order below $T_{\rm N} \sim$ 11~K.
  In the paramagnetic state above $T_{\rm N}$, from the analysis of NMR shift $K(T)$ using the high-temperature series expansion, $J/k_{\rm B}$ was estimated to be $ \simeq 1.9$~K, which is consistent with the previously reported value of $J/k_{\rm B} \simeq 1.8$~K.
   From the temperature dependences of 1/$T_1$ and 1/$T_2$ in the paramagnetic state above $T_{\rm N}$, AFM spin fluctuations were found to be developed below $\sim$ 50 K.
   In the AFM ordered state, we observed a rectangular shape of the spectrum, evidencing a commensurate AFM state.
   However, additional peaks in the spectra in the AFM state were observed above 3 T, which are probably due to the spin flop reported previously~\cite{Sebin}.
   The temperature dependences of the sublattice magnetization and 1/$T_1$ in the AFM state were well explained by spin waves  for a conventional three-dimensional antiferromagnet with a finite magnetic anisotropy gap $\sim$ 5.7 K, indicating that the AFM state is rigid without any obvious effects of spin frustrations in its ground state.

\section{Experimental }
   Polycrystalline Na$_3$Fe(PO$_4)_2$ sample was synthesized using the conventional solid-state method~\cite{Sebin}.  
   NMR measurements of $^{31}$P ($I$ = $ \frac{1}{2}$, $\frac{\gamma_N}{2\pi}$ = $17.237$ MHz/T) nuclei were conducted using a laboratory-built phase-coherent spin-echo pulse spectrometer.
   The $^{31}$P NMR spectra were obtained by sweeping magnetic field $H$ at fixed resonance frequencies.
   $^{31}$P zero-shift position ($H_{\rm ref}$) for each resonance frequency was determined by the $^{31}$P NMR of the nonmagnetic reference sample H$_3$PO$_4$.   
   The $^{31}$P spin-lattice relaxation rates $1/T_1$ were measured by the saturation recovery method.
$1/T_1$ at each temperature was determined by fitting the nuclear magnetization $M(t)$ versus time $t$ using the single exponential function
\begin{equation}
1-\frac{M(t)}{M(\infty)}= A~exp\left[ -\left(\frac{t}{T_1}\right)\right], 
\label{eqn:T1fit}
\end{equation}
where $M(\infty)$ is the equilibrium nuclear magnetization at $t \rightarrow \infty$.

Nuclear spin-spin relaxation times $T_2$ were determined by fitting spin-echo decay curves using  the following equation:
\begin{equation}
M(2\tau)= M_0~exp\left[ -\left(\frac{2\tau}{T_2}\right)^\beta\right], 
\label{eqn:T2fit}
\end{equation}
where $\tau $ is the time between $\pi$/2 and $\pi$ pulses.
In the paramagnetic state, Lorentzian relaxation behavior with $\beta = 1$ was observed.
However, in the AFM state, $\beta$ increases with decreasing temperature and becomes $\approx 2$ at low temperatures below $T = 3.4$~K, showing Gaussian decay behavior.

\section{Results and discussions}

\subsection{$^{31}$P NMR spectrum in the paramagnetic state}

\begin{figure}[tb]
  \includegraphics[width=12 cm]{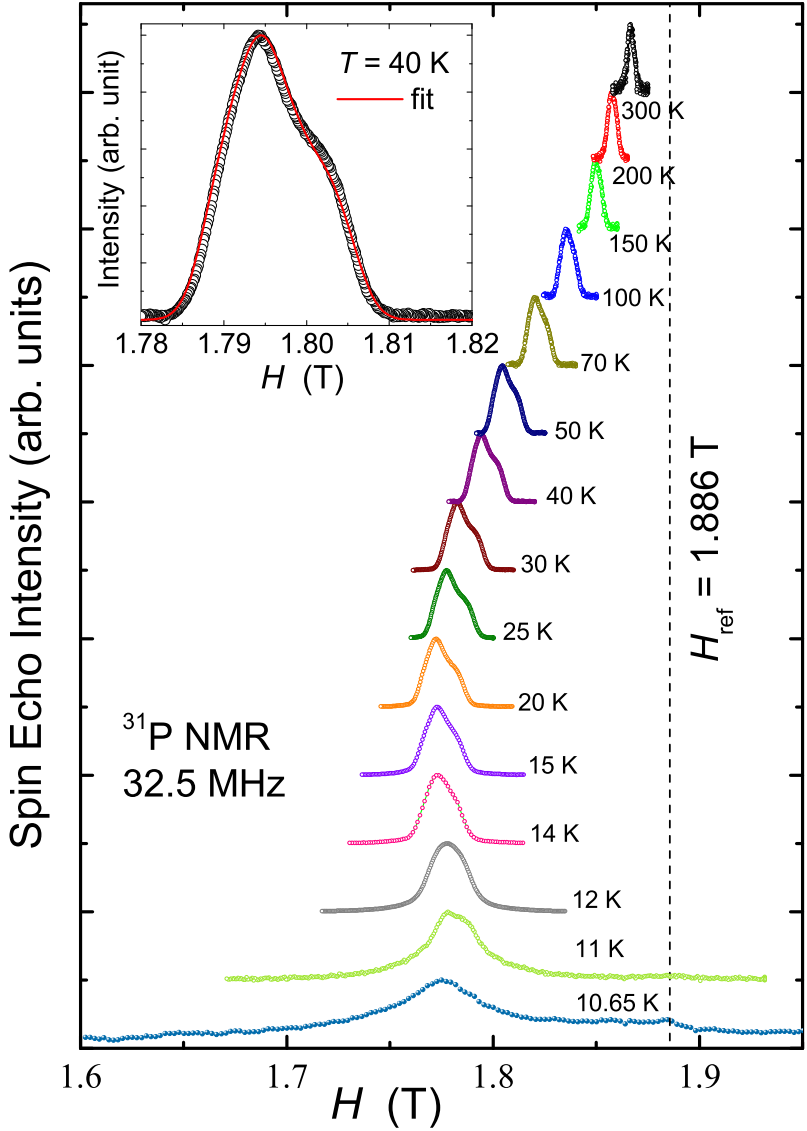}
\caption{ Field-swept $^{31}$P NMR spectra at different temperatures measured at $f$ = 32.5~MHz. The vertical dash-dotted line corresponds to the $^{31}$P zero-shift or non-magnetic reference field ($H_{\rm ref} \approx 1.886$~T). The inset shows the spectrum at 40~K with the simulated spectrum (red curve) with $K_{\rm iso} \simeq 4.91$\%, $K_{\rm ax} \simeq -0.28$\%.} 
\label{fig:spectrum_1}
\end{figure}

\begin{figure}[tb]
  \includegraphics[width=12 cm]{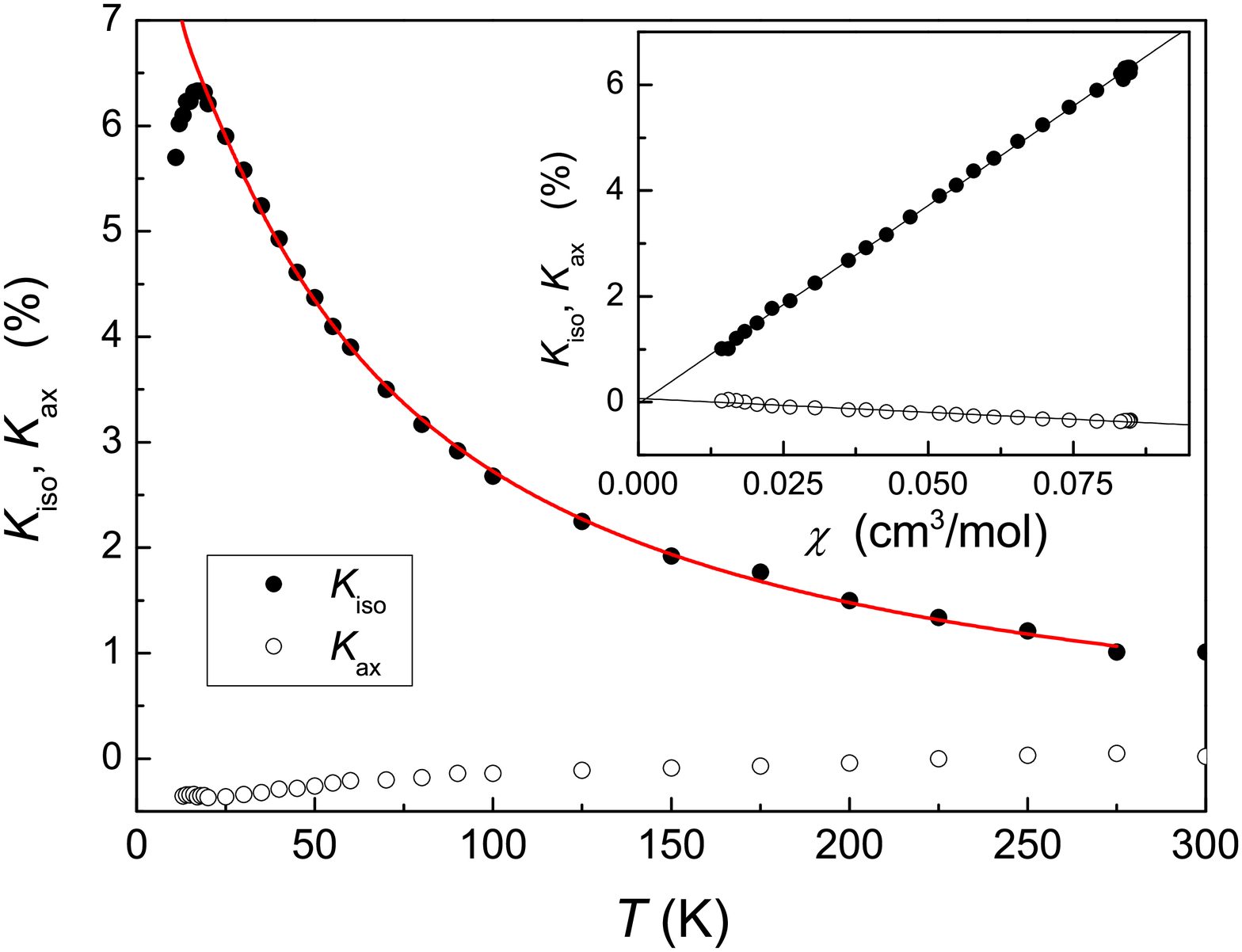}
	 \caption{Temperature-dependent $^{31}$P NMR iso-shift $K_{\rm iso}$ and $K_{\rm ax}$ measured at 32.5~MHz. 
       The red curve shows the result of the fit to  $K_{\rm iso}$  using Eq.~\ref{Eq2}, above 20~K. 
      Inset: $K$ vs. $\chi$ plots for  $K_{\rm iso}$ and  $K_{\rm ax}$ where we used $\chi$ data reported in Ref. \cite{Sebin}.
The thick solid lines are fitting results.  }
	\label{fig:NMR_shift}
\end{figure}

     Figure~\ref{fig:spectrum_1} shows the temperature dependence of the field-swept $^{31}$P NMR spectra measured on the polycrystalline sample at a frequency of \textit{f} = 32.5~MHz ($H_{\rm ref}\approx 1.886$~T) at several temperatures ($T>T_{\rm N}$).
The observed spectra show a characteristic shape of the anisotropic powder pattern.
With decreasing temperature, the spectra become broader and shift to lower magnetic fields. 
   Then the spectra move to slightly higher magnetic fields below 20 K.

     The asymmetric shape of spectra was well reproduced by the calculated spectra where we introduced an anisotropy in NMR shift $K$.
The red curve in the inset of Fig.~\ref{fig:spectrum_1} shows the typical calculated powder-pattern spectrum with isotropic and axial components of NMR shifts ($K_{\rm iso}$ and $K_{\rm ax}$) and reproduced the observed spectrum (open black circles) very well. 
 Here the NMR shift $K$ is described by $K$ = $K_{\rm iso}$ + $K_{\rm ax}$(3cos$^2$$\theta$ -1) where $\theta$  is the angle between the principal axis of hyperfine tensor at the P site and the external magnetic field~\cite{MetallicShifts}.


      From the fitting of the spectra measured at different temperatures, we determined the temperature dependences of $K_{\rm iso}$ and $K_{\rm ax}$ as shown in Fig. \ref{fig:NMR_shift}. 
     The temperature dependences of $K_{\rm iso}$ and $K_{\rm ax}$ are essentially the same physically, although their sign and magnitude are different. 
   The absolute values of $K_{\rm iso}$ and $K_{\rm ax}$ increase with decreasing temperature in the high-temperature region following the Curie-Weiss behavior. Further decrease in temperature leads to a broad maximum (minimum)  in  $K_{\rm iso}$ ($K_{\rm ax}$)  around 17~K, analogous to the $\chi({T})$ data  depicting a short-range AFM  ordering. 
    $K$ has a direct relation with the spin susceptibility  $\chi_{\rm spin}$, and is given by the equation
\begin{equation}
K = K_{0}+\frac{A_{\rm hf}}{N_{\rm A}\mu_{\rm B}}\chi_{\rm spin},
\label{Eq2}
\end{equation}
where $K_{0}$ is the temperature-independent chemical shift, $N_{A}$ is the Avogadro number, and $A_{\rm hf}$ is the hyperfine coupling constant between the $^{31}$P nuclei and the Fe$^{3+}$ electronic spins.
Thus, the slope from the linear fit of $K$ vs. $\chi$ plot with temperature as an implicit parameter yields the corresponding $A_{\rm hf}$.
    As shown in the inset of Fig.~\ref{fig:NMR_shift}, the linear dependence of $K_{\rm iso}$ and $K_{\rm ax}$ on the $\chi$ over the whole temperature range ($T>T_{\rm N}$)  can be seen where the $\chi$ data are obtained from Ref.~\cite{Sebin}, showing again the Curies-Weiss behavior in $K_{\rm iso}$ and $K_{\rm ax}$  in the temperature region.
 The slopes yield the isotropic part of the hyperfine coupling constant $A^{\rm iso}_{\rm hf}~\simeq (4.29 \pm 0.02) $~kOe/$\mu_{\rm B}$ and the axial part of the hyperfine coupling constant $A^{\rm ax}_{\rm hf}~\simeq (-0.29 \pm 0.01) $~kOe/$\mu_{\rm B}$.

The isotropic part of the $A_{\rm hf}$ originates from the transferred hyperfine field due to the P($3s$)-O($2p$)-Fe(3$d$) covalent bond.
On the other hand, the axial part of the $A_{\rm hf}$ could be due to an anisotropic hyperfine field at P atoms though the P($3p$)-O($2p$)-Fe(3$d$) covalent bond and/or classical dipolar field from Fe$^{3+}$ ions.
From our calculation for the classical dipolar field, it is found that the axial component of the classical dipolar field is $|$0.18$|$ kOe/$\mu_{\rm B}$, the same order as the observed $A^{\rm ax}_{\rm hf}~\simeq (-0.29 \pm 0.01) $~kOe/$\mu_{\rm B}$, indicating that the $A^{\rm ax}_{\rm hf}$ mainly comes from the classical dipolar field.

To estimate the exchange coupling $J$ between Fe$^{3+}$ spins, we fitted the temperature dependence of $K_{\rm iso}$ above 20 K by Eq.~\ref{Eq2}, where we used the high-temperature series-expansion of $\chi_{\rm spin}$ for a spin-$S$ Heisenberg isotropic triangle lattice model~\cite{Delmas55} given by,
\begin{equation}
\frac{N_{A}\mu_{B}^{2}g^{2}}{3 \left| J\right|  \chi_{\rm spin}}= x + 4+\frac{3.20}{x}-\frac{2.186}{x^2}+\frac{0.08}{x^3}+\frac{3.45}{x^4}-\frac{3.99}{x^5},
\label{Eq3}
\end{equation}
with $x= k_{\rm B}T/|J | S(S+1)$. The Eq.~\ref{Eq3} predicts the $\chi_{\rm spin}$ of a triangular lattice accurately for $T\geq(J/k_{\rm B})S(S+1)$~\cite{Schmidt104443}. 
   With using  $A^{\rm iso}_{\rm hf}~\simeq 4.29$~kOe/$\mu_{\rm B}$, we fitted the data as shown by the red  curve in  Fig. \ref{fig:NMR_shift} and estimated  $J/k_{\rm B}\simeq 1.9$~K, and Land\`{e} g-factor $g~\simeq 2.18$. 
  The exchange coupling obtained by this fit is close to the previously reported value of $J/k_{\rm B}$ = 1.8 K from $\chi(T)$ analysis~\cite{Sebin}.

\subsection{$^{31}$P NMR spectrum in the AFM state}
\begin{figure}
\includegraphics[width=10 cm] {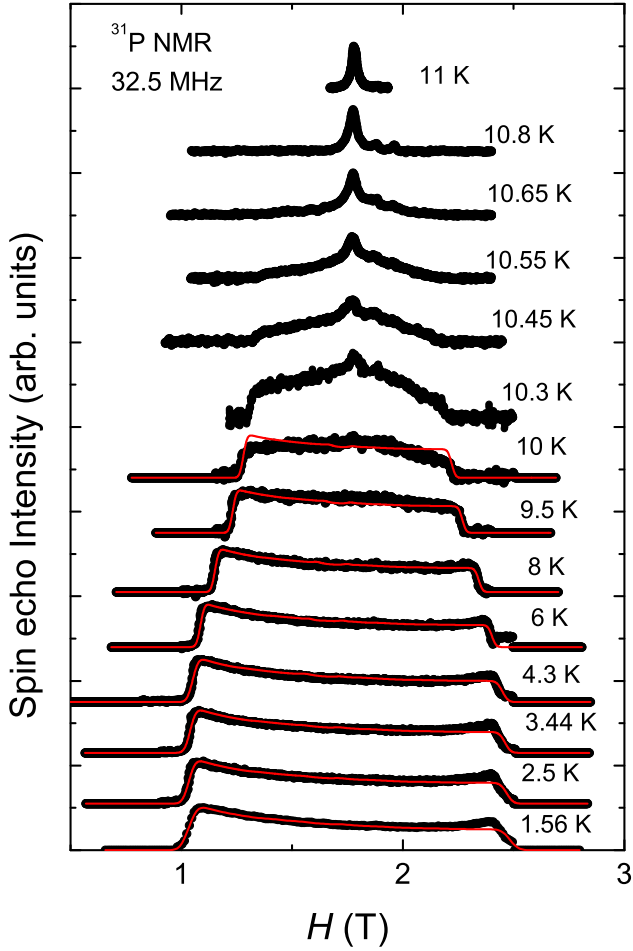}
\caption{\label{fig:spectrum_2}Temperature-dependent $^{31}$P NMR spectra measured at 32.5~MHz, below $T_{\rm N}$. The $H_{\rm int}$ broadens the spectra below $T_{\rm N}$, taking a rectangular shape. The red curves show the calculated spectra using Eq.~\ref{Eq7} for a commensurate AFM ordered state.}
\end{figure}

The $^{31}$P  spectra measured at 32.5~MHz in the AFM state are shown in Fig.~\ref{fig:spectrum_2}. 
The spectra  broaden below $T_{\rm N}$ and exhibit nearly  rectangular shapes at low temperatures. 
The immediate line broadening below $T_{\rm N}$ is due to the static internal field at the P site produced by the ordered moments of Fe$^{3+}$ ions in the magnetically ordered state. 

  In general, rectangular-shape NMR spectra are expected in rigid antiferromagnets in powder samples where the angle between the internal field $H_{\rm int}$ and the external magnetic field is randomly distributed \cite{Yamada}.
  These rectangular spectra were observed in several other compounds showing a commensurate AFM ordered state~\cite{Ranjith024422}. 

   Assuming a uniform  $H_{\rm int}$ in a commensurate AFM ordered state, 
one can express the NMR spectra, $f(H)$, as ~\cite{Ranjith014415,Nath024431}
\begin{equation}
f(H)\propto \frac{H^2-H^2_{\rm int}+\omega^2_{\rm N}/\gamma^2_{\rm N}}{H_{\rm int}H^2},
\label{Eq6}
\end{equation}
where $\omega_{\rm N}/\gamma_{\rm N} = H_0$  is the central field, $\omega_{\rm N}$ is the NMR frequency, $\gamma_{\rm N}$ is the nuclear gyromagnetic ratio, and $H_{\rm int}$ is the internal field. 
The cut-off fields are represented as $H_0 + H_{\rm int}$ and $H_0 - H_{\rm int}$, which produces the two sharp edges of the spectrum.
Due to the small distribution of internal fields, the internal field $H_{\rm int}$ is modeled by a Gaussian distribution function to smoothen the sharp edges. 
  To obtain the spectra, the function $f(H)$ and the distribution function $g(H)$ were convoluted as
\begin{equation}
F(H) =  \int_{0}^{\infty} f(H-H')g(H') \,dH',
\label{Eq7}
\end{equation}
where the Gaussian distribution function is given by 
\begin{equation}
g(H)=\frac{1}{\sqrt{2\pi\Delta H^2_{\rm int}}}exp\left(-\frac{(H-H_{\rm int})^2}{2\Delta H^2_{\rm int}}\right).	
\label{Eq8}
\end{equation}
The red curves in Fig.~\ref{fig:spectrum_2} show calculated spectra, which reproduce the observed spectra well.
This indicates a commensurate AFM magnetic order below $T_{\rm N}$, consistent with neutron diffraction measurements~\cite{Sebin}.
    It is noted that there is a small deviation between the calculated and observed spectra around the edge of the spectrum at the high magnetic field side. 
    It is probably due to the additional signal intensity due to $^{23}$Na NMR signals (not shown) and/or it may be due to partial orientation of the powder samples originating from the application of magnetic field. 

\begin{figure}
\includegraphics[width=13 cm] {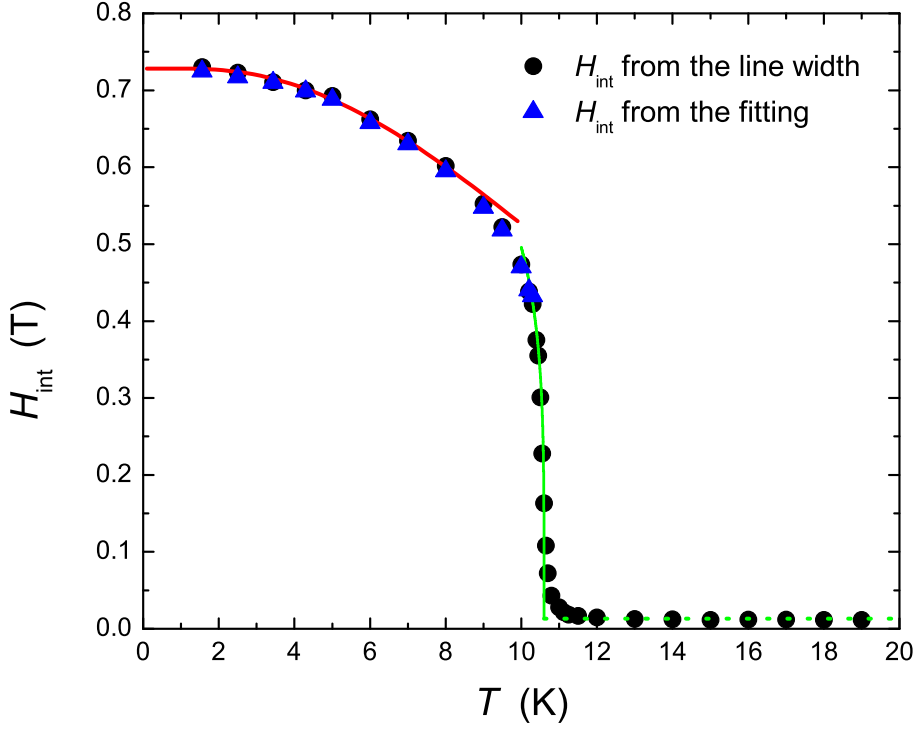}
\caption{\label{fig:Bint} Temperature dependence of the internal field $H_{\rm int}$ obtained from the NMR spectra in the AFM state.
The blue triangles and black closed circles show the $H_{\rm int}$  estimated from fittings and from the half of full-width at 25 \% of the peak intensity of spectra, respectively. 
The green solid curve is the fit of $H_{\rm int}$ vs. $T$ using Eq.~\ref{Eq9} in the critical regime, and the green dotted line exhibits the constant term of Eq.~\ref{Eq9} in the paramagnetic state. The red solid curve shows the calculated result using Eq.~\ref{eq:Sublatice} for the magnon model in the AFM state.
}
\end{figure}

     Figure~\ref{fig:Bint} shows the temperature dependence of $H_{\rm int}$ determined by the aforementioned fittings of the spectra (blue triangles). 
     It was difficult to precisely determine the $H_{\rm int}$ near $T_{\rm N}$ by the fittings because the commensurate AFM state is not well developed. 
    Therefore we also estimated  the $H_{\rm int}$ by taking half of full-width at $25 \%$ of the peak intensity of the spectra shown by the black closed circles  in order to determine $H_{\rm int}$ near $T_{\rm N}$. 
    As seen, both methods produce almost the same $H_{\rm int}$  at $T$ $<<$ $T_{\rm N}$.  
As we approach the $T_{\rm N}$, $H_{\rm int}$ decreases sharply. 
Since $H_{\rm int}$ is proportional to the Fe$^{3+}$ sublattice magnetization, $H_{\rm int}(T)$ close to $T_{\rm N}$ is related to the critical exponent ($\beta$) of the order parameter.  
  To estimate the value of $\beta$, we fit the $H_{\rm int}$ obtained by the line width in the critical regime, very close to $T_{\rm N}$ using the following equation
\begin{equation}
H_{\rm int}(T) = H_{\rm int}(0) \left(1-\frac{T}{T_N} \right)^{\beta} + c ,
\label{Eq9}
\end{equation}
where the constant value of $c$ is included to consider the line width in the paramagnetic state. 
 Our fit in the temperature range 10-10.6~K  by Eq.~\ref{Eq9} yields $\beta \simeq 0.29$ with $H_{\rm int}(0) \simeq 1.1$~T,   $T_{\rm N} = 10.6$~K and  $c \simeq 0.013$~T. 
 The obtained value of $\beta$ is slightly greater than  the expected value for 2D XY model ($\beta \simeq 0.231$)~\cite{Bramwell} and 2D Ising model ($\beta \simeq 0.25$) \cite{Collins,Ozeki} and is close to or is slightly smaller than the expected values for 3D system (0.31-0.326 for 3D Ising model, 0.31-0.345 for 3D XY model and 0.33-0.367 for 3D Heisenberg model \cite{Nath214430}).   
   On the other hand, when we change the fitting range to  9-10.6  K from 10-10.6  K, the $\beta$ value decreases to 0.23 which is close to the value for the 2D XY model.  
  Since $\beta$ = 0.29 is obtained in the temperature regime close to $T_{\rm N}$, our results may suggest that the spin dimensionality of  the system very close to $T_{\rm N}$  is most likely characterized by the 3D nature  in NaFe$_3$(PO$_4$)$_2$.

\begin{figure}
\includegraphics[width=15 cm] {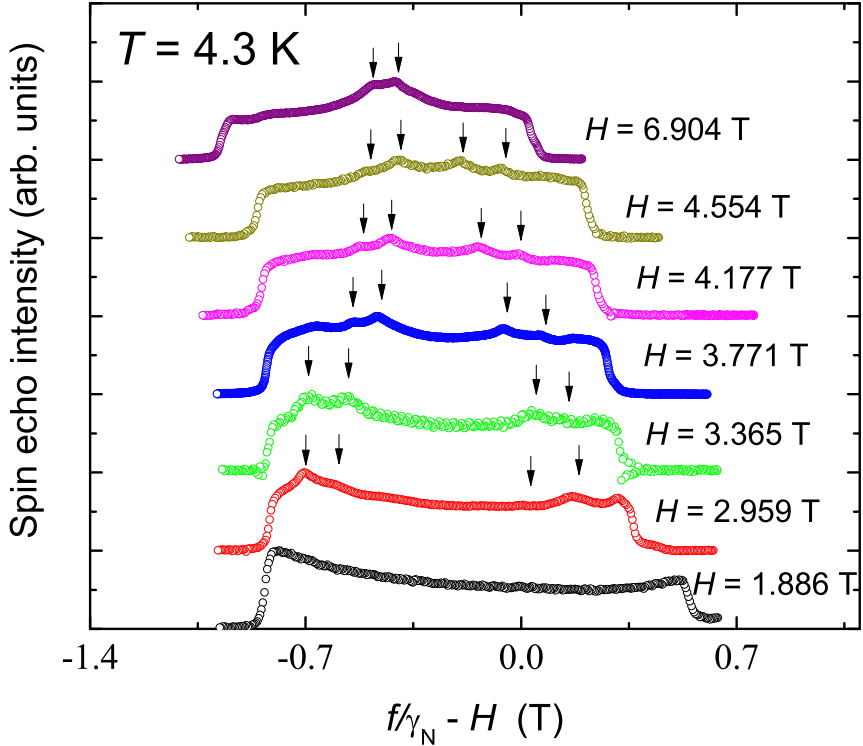}
\caption{\label{fig:H-spectrum} $^{31}$P NMR spectra measured at $T= 4.3$~K for different applied magnetic fields $H$, above and below the spin-flop transition. The arrows depict the evolution of extra features with the field above the spin-flop transition.}
\end{figure}

Finally, we show the  field-dependent $^{31}$P NMR spectra measured at $T= 4.3$~K  in Fig.~\ref{fig:H-spectrum}. 
Interestingly additional peaks were observed with the increase in $H$  above $\sim 3$~T, whose positions seem to move to the center of the spectra with  increasing $H$.
These results indicate the appearance of the P site with different $H_{\rm int}$ under magnetic fields.
It is interesting to point out that a spin flop has been reported around 3.2~T at 4.2~K from the magnetization measurements \cite{Sebin}.
Although the origin of those additional peaks in the NMR spectrum is not clear at present, if it were due to the spin flop, our experimental results suggest that the spin flop may take place partially, not uniformly, since the observed spectra consist of the rectangular powder pattern due to the uniform internal field and the additional peaks with different internal fields.
Further experiments such as neutron diffraction measurements under magnetic field and/or NMR measurements on single crystals are required to clarify this point.


\subsection{$^{31}$P spin-lattice relaxation rate in the paramagnetic state }

\begin{figure}
\includegraphics[width=10 cm] {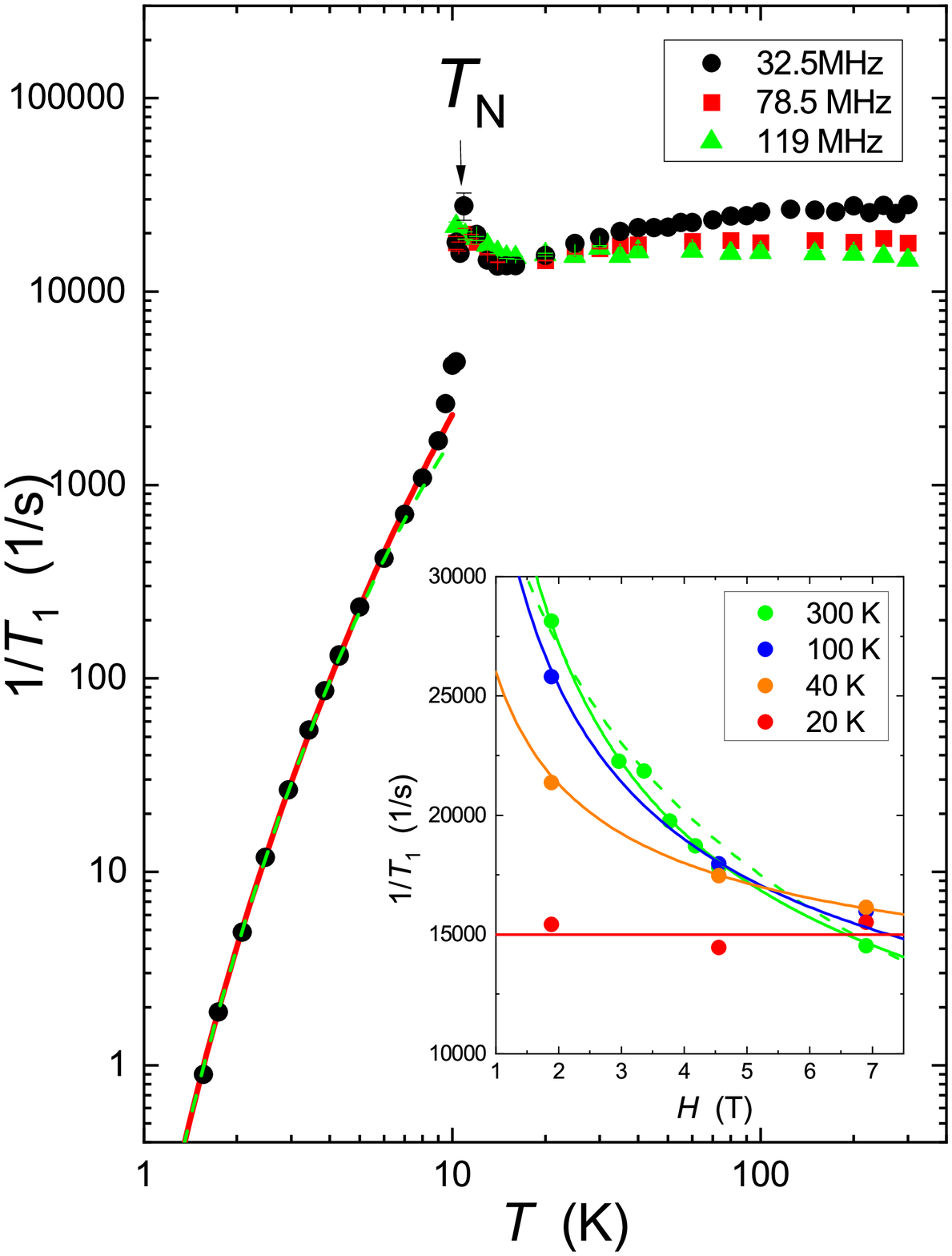}
\caption{\label{fig:T1} Temperature dependence of $1/T_1$ measured at three frequencies (32.5, 78.5, and 119~MHz).
The red solid line represents the fits using Eq. \ref{eq:T1_magnon} based on the model of the two magnon Raman process with $C$ = 2.5 (1/sK$^3$) and $T_{\rm AE}$ = 5.7 K in the AFM state. The green dotted line shows the calculated result with the model for triangular antiferromagnet using Eq. \ref{eq:T1_magnon2} with $C'$ = 1.3$\times$10$^{-3}$ (1/sK$^5$) and $T_{\rm AE}$ = 5.7 K.
Inset shows the $H$ dependence of $1/T_1$ at several temperatures (20, 40, 100, and 300 K).
The solid lines except for 20 K data are the expected $H$ dependence of 1/$T_1$ based on the 1D spin-diffusion model. 
It is noted that   1/$T_1$ data at 20 K shows a nearly $H$ independent behavior, indicating no obvious spin diffusion effects.
The green dotted curve shows the log(1/$H$) dependence of 1/$T_1$ expected for the 2D spin-diffusion model fitted the data at 300 K. }
\end{figure}

     The spin-lattice relaxation rate $1/T_{1}$ for $^{31}$P nuclei was measured at the peak position of the spectrum to understand the dynamical properties of Fe$^{3+}$ spins. 
Figure~\ref{fig:T1} shows the temperature dependence of $1/T_{1}$ measured at $f$ = 32.5 MHz.
Above $\sim$50~K, $1/T_{1}$ is nearly independent of temperature and can be explained by the random fluctuation of paramagnetic moments in the limit $T \gg J/k_{\rm B}$~\cite{Sebastian064413}.
As the temperature is lowered below $T$ $<$ $\sim$50~K, $1/T_{1}$ slowly decreases, showing a local minimum around 15 K, and then peaks at $T_{\rm N}\simeq10.9$~K, due to the magnetic long-range order.
Below $T_{\rm N}$, the $1/T_{1}$ decreases strongly with decreasing temperature.



In the paramagnetic state, 1/$T_1$ depends on the resonance frequency (magnetic field), as shown in Fig.~\ref{fig:T1}.
    1/$T_1$ decreases with an increase in frequency in the high-temperature region.
The difference in 1/$T_1$ is narrowed down as the temperature is lowered, and below $T \sim$ 20~K, the data sets in different frequencies nearly overlap with each other.
It is known that the long-wavelength ($q$ = 0) spin fluctuations in a Heisenberg magnet often show diffusive dynamics.
This is due to the divergence behavior of the spectral density of the spin-spin correlations as $\omega$ goes to zero, which depends on the dimensionality of the spin system.
In 1D spin chains, such a spin diffusion leads to a $1/\sqrt{H}$ field dependence of 1/$T_ 1$ as observed in Sr$_2$CuO$_3$~\cite{Takigawa4612} and (CH$_3$)$_4$NMnCl$_3$~\cite{Hone965}.
In the case of 2D materials, $1/T_{1}$ is known to vary as log$(1/H)$~\cite{Ajiro420,Furukawa2393,Yogi024413}.
In the inset of Fig.~\ref{fig:T1}, we plot $1/T_{1}$ against $H$ at several temperatures.
We fitted the experimental data at 300 K with the relations 1/$T_1$ $\propto$ $1/\sqrt{H}$ and 1/$T_1$ $\propto$ $\log{1/H}$ which are shown by the green solid and dotted curves in the inset of Fig.~\ref{fig:T1}.
As both $H$ dependences of $1/T_1$ capture the experimental observations, we do not conclude whether the system is close to the 1D or 2D system within our experimental uncertainty.
For other temperatures such as 40 K and 100 K, we tentatively show the fits for the 1D model.
It is noted that 1/$T_1$ data at 20 K shows a nearly $H$-independent behavior, indicating no obvious spin diffusion effects.
In any case, we consider that the $H$-dependent behavior of 1/$T_1$ observed at high temperatures in NaFe$_3$(PO$_4$)$_2$ originates from the spin diffusion effect, a characteristic of low-dimensional spin systems.


\begin{figure}
\includegraphics[width=13 cm] {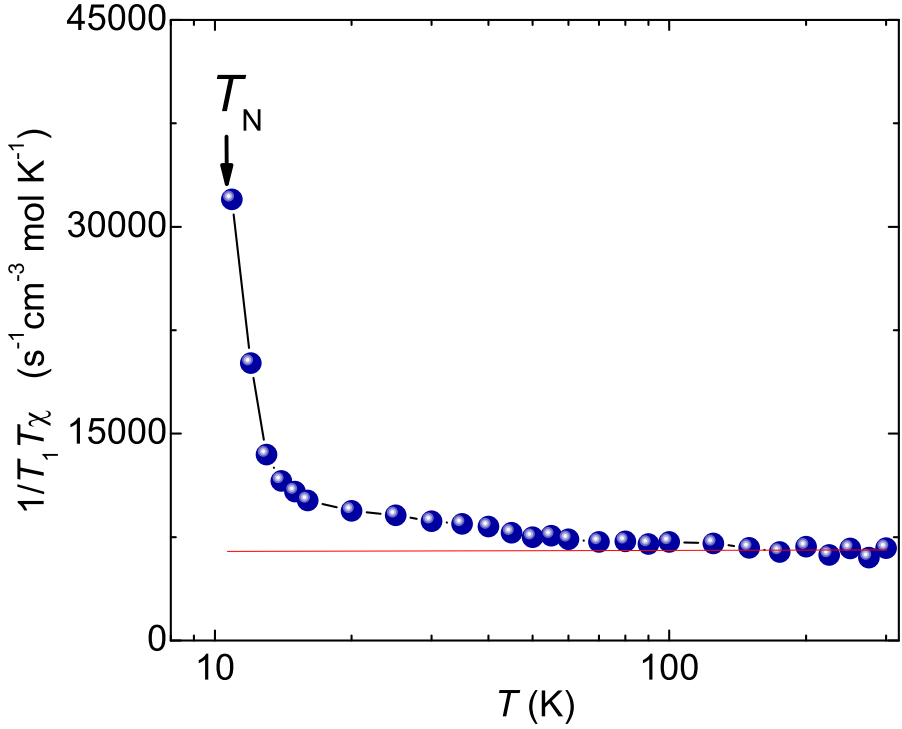}
\caption{\label{fig:T1Tchi} Temperature dependence of $1/T_{1} T \chi$ measured at $f$  = 32.5 MHz.}
\end{figure}

Finally, to see spin fluctuation effects  in the paramagnetic state above $T_{\rm N}$, it is useful to replot the data by changing the vertical axis from 1/$T_1$ to   $1/T_{1}T \chi$  as shown in Fig.~\ref{fig:T1Tchi}.
$1/T_{1}T$ is related to the imaginary part of dynamic susceptibility $\chi^{\prime \prime}\left(q, \omega_{\rm n}\right)$ at the NMR  frequency $\omega_{\rm N}$ as \cite{Smerald}
\begin{equation}
\frac{1}{T_{1} T}=\frac{2 \gamma_{N}^{2} k_{\mathrm{B}}}{N_{\mathrm{A}}^{2}} \sum_{q}|A(q)|^{2} \frac{\chi^{\prime \prime}\left(q, \omega_{\rm N}\right)}{\omega_{\rm N}},
\label{Eq4}
\end{equation}
where $k_{\mathrm{B}}$ is the Boltzmann’s constant. 
   In Eq.~\ref{Eq4}, the sum is over the wave vectors $\vec{q}$ within the first Brillouin zone, and $A(\vec{q})$ is the form factor of the hyperfine interactions. The uniform static susceptibility $\chi = \chi^\prime(0,0)$, is the real part $\chi^{\prime }\left(q, \omega_{\rm N}\right)$ with $q=0$ and $\omega_{\rm N} = 0$. Thus, by plotting  $1/T_{1}T \chi$ vs. $T$, we get the temperature dependence of $\sum_{q}|A(q)|^{2} \chi^{\prime \prime}\left(q, \omega_{\rm N}\right)$ compared to that of $\chi^\prime(0,0)$, the uniform susceptibility.

   As shown in Fig.~\ref{fig:T1Tchi}, $1/T_{1}T \chi$ is nearly independent of temperature in the high temperature region down to $T \sim 50$~K. 
   Thus, in this temperature region, $\sum_{q}|A(q)|^{2} \chi^{\prime \prime}\left(q, \omega_{\rm N}\right)$ scales to $\chi^\prime(0,0)$.  
    Below $T \sim $ 50~K till $T_{\rm N}$, it increases with decreasing temperature, implying that $\sum_{q}|A(q)|^{2} \chi^{\prime \prime}\left(q, \omega_{\rm N}\right)$ increases more than $\chi^\prime(0,0)$.  
This result evidences a growth of spin fluctuations with $q\neq 0$, attributing to AFM correlations.

\subsection{$^{31}$P spin-lattice relaxation rate in the AFM state }

 We observe a strong temperature dependence of spin-lattice relaxation rate, indicative of the nuclear relaxation due to scattering of magnons~\cite{Belesi184408}.
According to Beeman and Pincus \cite{Beeman359}, in the AF state for magnetic insulators, 1/$T_1$ is mainly driven by magnon processes, leading to a strong temperature dependence of 1/$T_1$.
Here, we calculate the temperature dependence of 1/$T_1$ based on the  two-magnon (Raman) process.
With a long wave approximation for magnon dispersion relation, 1/$T_1$ can be given by \cite{Beeman359}
\begin{equation}
\frac{1}{T_1} = C T^3  \int_{T_{AE}/T}^{\infty} \frac{x}{e^x-1}dx.
\label{eq:T1_magnon}
\end{equation}
Here, the fitting parameters are the proportional constant $C$ and $T_{\rm AE}$. 
  $T_{\rm AE}$ is the anisotropy gap in the spin wave spectrum.
  The solid red curve  in Fig. \ref{fig:T1} shows the calculated results with  $C$ = 2.5 (1/sK$^3$) and $T_{\rm AE}$ = 5.7 K, which reproduced the experimental data well.

   We also calculated the temperature dependence of 1/$T_1$ for the two magnon process discussed in triangular antiferromagnets CsNiBr$_3$ \cite{Maegawa1995} and NiGa$_2$S$_4$ \cite{Takeya2008}, and a Kagome lattice antiferromagnet KFe$_3$(OH)$_6$(SO$_4$)$_2$ \cite{Nishimiya2003}.
In this case, the temperature dependence of 1/$T_1$ is given by 
\begin{equation}
\frac{1}{T_1} = C\rq{} T^5  \int_{T_{\rm AE}/T}^{T_{\rm m}/T} \left[x^2-\left(\frac{T_{\rm AE}}{T}\right)^2\right] \left[x^2+\left(\frac{T_m}{T}\right)^2\right] \frac{e^x}{(e^x-1)^2}dx,
\label{eq:T1_magnon2}
\end{equation}
where $T_m$ is defined by $ \hbar \omega_{\rm m} /k_{\rm B}$ with the maximum frequency of the spin wave ($\omega_{\rm m}$), and $T_{\rm m} = zSJ$ ($\sim$ 29 K) where $z = 6$ is the number of nearest neighbor of Fe ions surrounding a Fe ion.
As shown by the green dotted curve, which is the calculated result with $C'$ = 1.3$\times$10$^{-3}$ (1/sK$^5$) and $T_{\rm AE}$ = 5.7 K, the model also reasonably reproduces experimental results.
Since both models reproduce the experimental data very well, the magnetic fluctuations in the AFM state of NaFe$_3$(PO$_4$)$_2$ are considered to be well described by the conventional magnon scattering without showing any additional magnetic fluctuations due to spin frustrations, if any.

The magnon picture also well reproduces the temperature dependence of the sublattice magnetization at low temperatures.
Using the long wave approximation, the temperature dependence of the sublattice magnetization is given by
\begin{equation}
M(0)-M(T) \propto T^2 \int_{T_{AE}/T}^{\infty} \frac{x}{e^x-1}dx .
\label{eq:Sublatice}
\end{equation}
The red curve in Fig. \ref{fig:Bint} shows the calculated results with the same value of the parameter ($T_{\rm AE}$ = 5.7 K), which can reproduce the temperature dependence of $H_{\rm int}$.
The result again shows the conventional nature of the AFM state in NaFe$_3$(PO$4$)$_2$.

Finally, we briefly mention magnetic anisotropy.
Experimentally the magnetic anisotropy in NaFe$_3$(PO$4$)$_2$ is clearly observed in the present $T_1$ data in the AFM state and also the magnetization measurements reported previously \cite{Sebin}.
Since the orbital moments of Fe$^{3+}$ ions are zero in the $^6S$ ground state for the 3$d^5$ electron configuration, one may expect no single-ion anisotropy in the first order.
However, because of the second-order perturbation between the ground and excited states, the single ion type anisotropy can be induced by spin-orbit interactions.
Dzyaloshinsky-Moriya interaction also may produce magnetic anisotropy in the magnetically ordered state.

\subsection{Nuclear spin-spin relaxation rate ($1/T_2$)}

\begin{figure}
\includegraphics[width=13 cm] {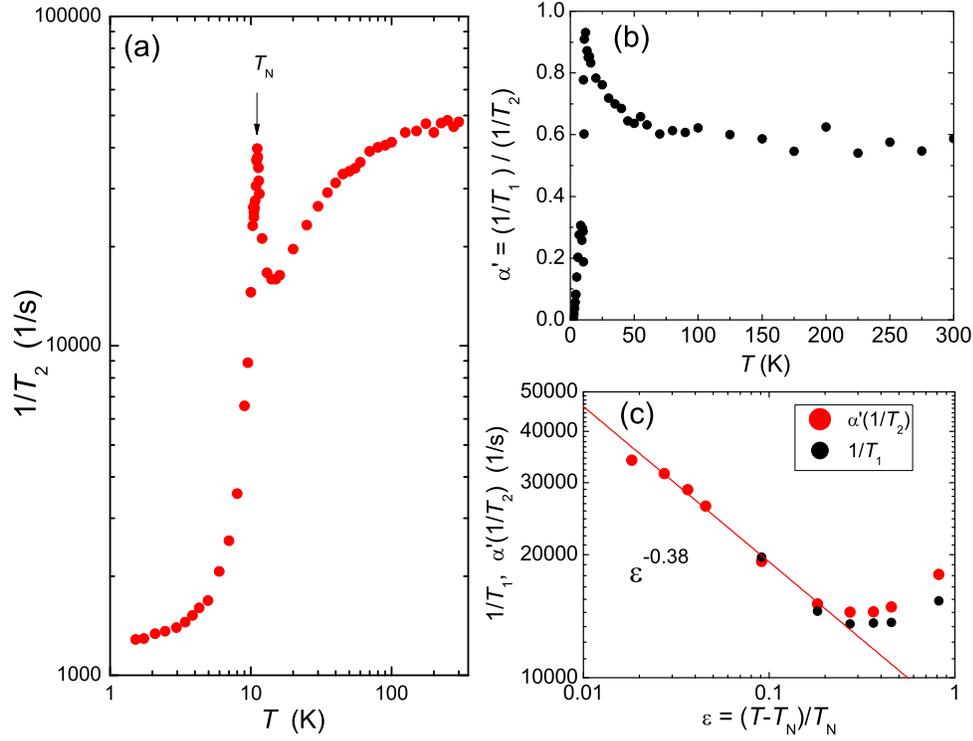}
\caption{\label{fig:T2}(a) $^{31}$P spin-spin relaxation rate $1/T_2$ vs. $T$ measured at 32.5~MHz. 
(b) Temperature dependence of $\alpha'$ [$\equiv$ (1/$T_1$)/(1/$T_2$)].
(c) 1/$T_1$ vs. $\varepsilon$   and  $\alpha\rq{}$/$T_2$ vs $\varepsilon$  where $\varepsilon$ = $(T-T_{\rm N})/T_{\rm N}$ with $T_{\rm N}$ = 11 K.
}
\end{figure}

The temperature dependence of $1/T_2$ measured at $f$ = 32.5 MHz is shown in Fig.~\ref{fig:T2}(a), and the trend agrees with that of $1/T_{1}$ behavior.
It can be seen that for $T\leq100$~K, $1/T_2$ gradually decreases down to $T\simeq15$~K.
Then 1/$T_2$ increases sharply and exhibits a peak at $T_{\rm N} \sim 11$~K, a clear indication of AFM long-range order.
Below $T_{\rm N}$, $1/T_2$ decreases and levels off at low temperatures.
It is noted that the values of 1/$T_2$ are slightly greater than those of $1/T_1$ measured at the same magnetic field above $T_{\rm N}$, although there is a large difference between them in the AFM ordered state.

In general, 1/$T_2$ is related to 1/$T_1$ and can be written for the case of $I$ = 1/2 as
\begin{equation}
\frac{1}{T_2 } = \left(\frac{1}{T_2}\right)^* +  \left(\frac{1}{2T_1}\right) +F_z(0),
\label{eqn:T2}
\end{equation}
with 
\begin{equation}
\frac{1}{T_1 } = F_\perp(\omega_{\rm N}). 
\label{eqn:T2-T1}
\end{equation}
Here, $F_\alpha(\omega)$ is the spectral density of the longitudinal ($\alpha$ = $z$) and transverse ($\alpha$ = $\perp$) components of the fluctuating local field ($h_{\alpha}$),  and is described by 
 \begin{eqnarray}
 F_\alpha (\omega_{\rm N})   
 =  \frac{1}{2} \gamma_{\rm N}^2 \int_{-\infty}^{+\infty} \langle \ h_{\alpha}(t) h_{\alpha}(0) \ \rangle  {\rm exp}(i \omega_{\rm N}t) dt.  
 \label{eqn:correlation}
\end{eqnarray}
$(1/T_2)^*$ is due to the nuclear dipole-dipole interaction \cite{Abragam, Slichter} and is independent of temperature.
This can be attributed to the nearly temperature-independent 1/$T_2$ observed at low temperatures in the AFM state.
The lowest value of 1/$T_2$ $\sim$ 1300 (1/s) corresponds to the spectral width $0.8$ Oe, which will be of the order of nuclear dipolar field at the P site.

Thus the temperature dependent part of 1/$T_2$ comes from the other two terms originating from $F_z(0)$ and $F_\perp(\omega_{\rm N})$.
To see how those contributions to 1/$T_2$ change with temperature, we plotted $\alpha'$ [$\equiv$ (1/$T_1$)/(1/$T_2$)] shown in Fig.~\ref{fig:T2}(b) where we used the 1/$T_1$ data measured at the same frequency of 32.5 MHz.
$\alpha'$ is nearly constant above 50 K, and starts increasing below 50 K with decreasing temperature and becomes close to $\alpha'$ $\sim$ 0.9 near $T_{\rm N}$.
Just below $T_{\rm N}$, $\alpha'$ shows a sudden decrease due to AFM ordering.
   Using $\alpha'$ = 0.6 above 50 K,   $F_z(0)$ is estimated to be 1.17$F_\perp(\omega_{\rm N})$. 
  This suggests that $F_z(0)$ and $F_\perp(\omega_{\rm N})$  contribute nearly equally to 1/$T_2$.  
Below 50 K, on the other hand, the increase in $\alpha'$ suggests that the contribution of $F_z(0)$ to 1/$T_2$ becomes small. 
   If we take   $\alpha'$ = 0.9,  $F_z(0)$ is calculated to be 0.61$F_\perp(\omega_{\rm N})$. 
   As we observed the development of AFM spin fluctuations below 50 K, the increase of $\alpha'$ could be due to the AFM spin fluctuations.
Since $F_\perp(\omega_{\rm N})$, that is, 1/$T_1$, picks up the hyperfine fluctuations at NMR frequency of $\omega_{\rm N}$ while $F_z(0)$ relates to the hyperfine fluctuations at $\omega$ = 0, the increase of $\alpha'$ may suggest that the spectrum density of the AFM spin fluctuation does not extend to $\omega$ = 0 in the paramagnetic state.

On the other hand, we observed the clear divergence behaviors in 1/$T_2$ and $1/T_1$ very near $T_{\rm N}$ which is due to the critical slowing down of magnetic fluctuations to $\omega$ = 0, characteristic of second order phase transition.
To see the critical behavior we plotted 1/$T_1$ and $\alpha'$/$T_2$ near $T_{\rm N}$ in Fig.~\ref{fig:T2}(c) as a function of $\varepsilon$ = $(T-T_{\rm N})/T_{\rm N}$.
The divergent behaviors are fitted with the power law of $1/T_1$ (or $\alpha$/$T_2$) $\propto$ $\varepsilon^{-0.38}$.
This value of power is close to 0.3 expected for the 3D Heisenberg spin system but far from 0.8 for the 2D Heisenberg spin system \cite{Benner Boucher}, indicating that the AFM transition is most likely characterized by the 3D nature.

\section{Conclusion}
The dynamic and static properties of the 2D anisotropic triangular lattice compound Na$_3$Fe(PO$_4)_2$ have been investigated from a microscopic point of view by the NMR technique.
The $^{31}$P  NMR spectra, nuclear spin-lattice relaxation rate 1/$T_1$ and nuclear spin-spin relaxation rate 1/$T_2$ results show the antiferromagnetic ground state below $T_{\rm N} \sim$ 11~K.
In the paramagnetic state above $T_{\rm N}$, from the analysis of NMR shift $K(T)$ using the high-temperature series expansion, $J/k_{\rm B}$ was estimated to be $ \simeq 1.9$~K, which is consistent with the previously reported value of $J/k_{\rm B} \simeq 1.8$~K.
The antiferromagnetic spin fluctuations were found to be developed below $\sim$50 K from the temperature dependence of $1/T_1T\chi$.
In the magnetically ordered state below $T_{\rm N}$, from the observation of the characteristic rectangular shape of the NMR spectra, the AFM state was found to be a commensurate AFM state in its ground state, consistent with the neutron diffraction measurements \cite{Sebin}.
The static and dynamic magnetic properties in the AFM state were revealed to be well described by the conventional spin-wave theory for a 3D antiferromagnet with a spin-anisotropy energy gap of 5.7 K.
Our NMR results indicate that the magnetically ordered state of Na$_3$Fe(PO$_4$)$_2$ is a conventional 3D antiferromagnetic state without any obvious effects of spin frustrations in its ground state.

 \section {Acknowledgments}
The research was supported by the U.S. Department of Energy (DOE), Office of Basic Energy Sciences, Division of Materials Sciences and Engineering. Ames Laboratory is operated for the U.S. DOE by Iowa State University under Contract No.~DE-AC02-07CH11358.  SJS and RN would like to acknowledge SERB, India, for financial support bearing sanction Grant No. CRG/2019/000960. SJS is supported by the Prime Minister's Research Fellowship (PMRF) scheme, Government of India.

\end{document}